\documentstyle[12pt]{article}
\newcommand{\be}{\begin{equation}}
\newcommand{\ee}{\end{equation}}
\newcommand{\bea}{\begin{eqnarray}}
\newcommand{\eea}{\end{eqnarray}}
\newcommand{\norsl}{\normalsize\sl}
\newcommand{\norsc}{\normalsize\sc}
\def \ksl {k \kern-.45em{/}}
\def \ppsl {p \kern-.45em{/}}
\def \nsl {n \kern-.45em{/}}

\begin{document}

\begin{titlepage}

\title{ Gauge-independent Thermal $\beta$ Function \\
                          In Yang-Mills Theory}
\author{
\norsc  Ken SASAKI\thanks{e-mail address: sasaki@mafalda.physics.nyu.edu
or sasaki@ed.ynu.ac.jp}\\
\norsl  Dept. of Physics, New York University\\
\norsl  4 Washington Place, New York, New York 10003, U.S.A.\\
\norsl  and \\
\norsl  Dept. of Physics, Yokohama National University\\
\norsl  Yokohama 240, JAPAN\\}

\date{}
\maketitle

\begin{abstract}
{\normalsize It is proposed to use the pinch technique (PT) to obtain
the gauge-independent thermal $\beta$ function in a hot Yang-Mills gas.
Calculations of the thermal $\beta$ function are
performed at one-loop level in four different gauges,
(i) the background field method with an arbitrary gauge,
(ii) the Feynman gauge, (iii) the Coulomb gauge, and (iv)
the temporal axial gauge, and they yield the same result in all four cases.}
\end{abstract}


\thispagestyle{empty}
\end{titlepage}
\setcounter{page}{1}
\baselineskip 18pt

It is important for the study of the quark-gluon plasma and/or
the evolution of the early Universe to fully understand the
behaviour of the effective coupling constant $\alpha_s (=g^2/{4\pi})$
in QCD at high temperature. The running of $\alpha_s$ with the
temperature $T$ and the external momentum $\kappa (=\vert \vec {\bf k} \vert)$
is governed by the thermal $\beta$ function $\beta_T$~\cite{Umezawa}. However,
the previous calculations of $\beta_T$ have exposed various
problems~\cite{EK}, a serious one of which is that the results
are gauge-fixing dependent~\cite{Niegawa}.

The background field method (BFM) has been applied to the calculation
of $\beta_T$ at one-loop~\cite{EK}\cite{BFM}\cite{ACPS}.
First introduced by DeWitt~\cite{Dea}, BFM is a technique for
quantizing gauge field theories while retaining explicit
gauge invariance for the background fields.
Since the Green's functions
constructed by BFM manifestly maintain gauge invariance, they
obey the naive QED-like Ward identities. As a result,
the spatial part of the three-gluon vertex, for static and
symmetric external momenta, is related to the transverse function
$\Pi_T(T,k_0=0,\kappa=\vert \vec {\bf k} \vert)$ of
the polarization tensor $\Pi_{\mu \nu}$, and thus $\beta_T$ is obtained
in BFM from~\cite{EK}
\be
   \beta_T \equiv T\frac{dg(T,\kappa)}{dT}=\frac{g}{2\kappa^2}
              T \frac{d\Pi_T (T,\kappa)}{dT}.
\label{Beta}
\ee
Due to the $O(3)$ invariance, the spatial
part of the gluon plarization tensor $\Pi_{ij}$ is expressed as follows:
\be
    \Pi_{ij}(k) = \Pi_T (\delta_{ij}-\frac{k_i k_j}{{\vec {\bf k}}^2}) +
                \Pi_L \frac{k_i k_j}{{\vec {\bf k}}^2}
\ee
and $\Pi_T$ can be extracted by
applying the projection operator
\be
           t_{ij}=\frac{1}{2}(\delta_{ij}-\frac{k_i k_j}{{\vec {\bf k}}^2})
\ee
to $\Pi_{ij}$.

The thermal $\beta$ function has been calculated in BFM at one-loop level
for the cases of the gauge parameter $\xi_Q=0$~\cite{BFM},
$\xi_Q=1$~\cite{ACPS} and
$\xi_Q=$an arbitrary number~\cite{EK}\cite{EHKT}. The results are
expressed in a form,
\be
    \beta_T^{BFM}=\frac{g^3 N}{128}\biggl\{
              \frac{7}{16}-\frac{1}{8}(1-\xi_{Q})+
                    \frac{1}{64}(1-\xi_{Q})^2 \biggr\}\frac{T}{\kappa},
\label{BetaBFM}
\ee
where $N$ is the number of colors.
Contrary to the case of the QCD $\beta$ function at zero temperature,
$\beta_T^{BFM}$ is dependent on the gauge-parameter $\xi_Q$.
The reason why we have obtained $\xi_Q$-dependent $\beta_T$ in BFM
is that the contributions to $\beta_T$ come from the finite part of
the gluon polarization tensor $\Pi_{\mu\nu}$ and that
BFM gives $\xi_Q$-dependent finite part for $\Pi_{\mu\nu}$
{}~\cite{Papavass}\cite{SirlinKostas}.

In this paper I propose to use the pinch technique (PT) to obtain the
gauge-independent $\beta_T$ in a hot Yang-Mills gas.
The PT was proposed some time ago by
Cornwall~\cite{rCa} for an algorithm to
form new gauge-independent proper vertices and new propagators
with gauge-independent self-energies.
First it was used to obtain the one-loop
gauge-independent effective gluon self-energy and vertices in
QCD~\cite{rCP} and then it has been
applied to the standard model~\cite{StandardModel}.

For example, let us consider the $S$-matrix element
$T$ for the elastic quark-quark scattering at one-loop order.
Besides the self-energy diagram in Fig.1,
the vertex diagrams of the first kind and
the second kind, and the box diagrams, which are shown in
Fig.2(a), Fig.3(a), and Fig.4(a), respectively, contribute to $T$.
Such contributions are, in general, gauge-dependent while the sum
is gauge-independent. Then
we can extract the ``pinch parts'' of the vertex and box diagrams,
which are dipicted in Fig.2(b), Fig.3(b), and Fig.4(b).
They emerge when a $\gamma^{\mu}$ matrix on the
quark line is
contracted with a four-momentum $k_{\mu}$ offered by a gluon propagator or an
elementary three-gluon vertex. Such a term triggers an elementary Ward
identity of the form
\be
      \ksl = ({\ppsl} + \ksl -m) - ({\ppsl} -m).
\label{Ward}
\ee
The first term removes (pinches out) the internal fermion propagator,
whereas the second term vanishes on shell, or {\it vice versa} . This leads to
contributions to $T$ with one less fermion propagator and, hence, these
contributions are called ``pinch parts''. The contribution of the
self-energy diagram, when added pinch parts from the vertex and box diagrams,
become gauge-independent. In this way we can construct the gauge-independent
effective gluon polarization tensor (self-energy).

As in the case of BFM, the effective gluon polarization tensor and vertices
constructed by PT obey the naive QED-like Ward identities. Thus we can use
the same Eq.(\ref{Beta}) to obtain $\beta_T$ in the framework of PT.
More importantly, PT gives the gauge-independent results {\it up to
the finite terms}, since they are constructed from $S$-matrix.
It was shown recently~\cite{Hiroshima}
that BFM with the gauge parameter $\xi_Q =1$ reproduces the PT results
at one-loop order.
However, for $\xi_Q \not= 1$, this coincidence does not hold any more.
In fact, BFM gives at one-loop order the gluon polarization tensor whose
finite part is $\xi_Q$-dependent. Interestingly enough,
Papavassiliou~\cite{Papavass} showed that when PT is applied to BFM
for $\xi_Q \not= 1$ to construct the effective gluon polarization tensor,
the gauge dependence of the finite part disappears and the
previous $\xi_Q =1$ result (or the universal PT result) is obtained.

To my knowledge, there exists, so far, only one approach, i.e. PT,
which gives the gauge-independent gluon polarization tensor
{\it up to finite terms}.
And these finite terms give contributions to $\beta_T$.
This notion inspires the use of PT for the calculations of $\beta_T$.
In the following I will show that we obtain
the same $\beta_T$ in the framework of PT
even when we calculate in four different gauges,
(i) the background field method with an arbitrary gauge,
(ii) the Feynman gauge, (iii) the Coulomb gauge, and (iv)
the temporal axial gauge.

\bigskip

\noindent
(i) {\it The Background Field Method}
\smallskip

In the background field method with an arbitrary gauge, the gluon propagator,
$iD_{ab(BFM)}^{\mu \nu}=-i\delta _{ab}D_{(BFM)}^{\mu \nu}$, and
the three-gluon vertex with one background gluon field,
$\widetilde{\Gamma}^{abc}_{\lambda \mu \nu }$, are given, respectively, as
follows~\cite{Abbott}:
\be
       D_{(BFM)}^{\mu \nu}=\frac{1}{k^2}
            \biggl[ g^{\mu \nu}-(1-\xi_Q)
       \frac{k^\mu k^\nu}{k^2}\biggr],
\label{propagator}
\ee
and
\be
     \widetilde{\Gamma}^{abc}_{\lambda \mu \nu } (p, k, q) = g f^{bac}
              \biggl[(1-\frac{1}{\xi_Q}) \Gamma^{P}_{\lambda \mu \nu}(p, k, q)
              + \Gamma^{F}_{\lambda \mu \nu}(p, k, q) \biggr],
\ee
where
\bea
     \Gamma^{P}_{\lambda \mu \nu}(p, k, q) &=&
               p_{\lambda}g_{\mu \nu} -q_{\nu} g_{\lambda \mu} \nonumber \\
     \Gamma^{F}_{\lambda \mu \nu}(p, k, q) &=&
            2k_{\lambda}g_{\mu \nu}-2k_{\nu}g_{\lambda \mu}
             -(2p+k)_{\mu}g_{\lambda \nu}.
\label{Vertex}
\eea
In the vertex, $k_\mu$ is taken to be the momentum of the
background field and each momentum flows inward and, thus, $p+k+q=0$.

A one-loop calculation of the polarization tensor was performed
in Ref.~\cite{EHKT}, from which the transverse function
$\Pi_{T}^{(BFG)}(k_0=0, \kappa=\vert \vec {\bf k} \vert)$
in the static limit can be
extracted for  $\kappa<<T$ as follows:
\be
     \Pi_{T}^{(BFM)}(T, \kappa)=Ng^{2}{\kappa}T \biggl\{
              \frac{7}{16}-\frac{1}{8}(1-\xi_{Q})+
                    \frac{1}{64}(1-\xi_{Q})^2 \biggr\} + {\cal O} (\kappa^2).
\label{BFM}
\ee
Using this expression for $\Pi_{T}$ in Eq.(\ref{Beta}), Elmfors and Kobes
obtained Eq.(\ref{BetaBFM}) for $\beta_T^{BFM}$
which is indeed gauge-parameter $\xi_Q$ dependent~\cite{EK}.

Now we evaluate the pinch contributions to $\Pi_{T}$.
We consider the quark-quark scattering at one-loop order in the
Minkowski space. We use the gluon propagator and the three-gluon
vertex given in Eqs.(\ref{propagator})-(\ref{Vertex}).
The pinch contributions come from
the vertex diagrams of the first kind [Fig. 2(b) and its mirror graph],
the vertex diagrams of the second kind [Fig. 3(b) and its mirror graph]
and the box-diagrams [Fig. 4(b)].
We can extract from them the pinch contribution to the polarization tensor,
which is expressed as~\cite{Papavass}
\bea
   i \Pi^{\mu \nu}_{P(BFM)} &=& N g^2 (1-\xi_{Q}) k^2
            \int \frac{d^4 p}{(2\pi )^4}
             \frac{-2kp}{p^4 q^2} g^{\mu \nu}  \nonumber  \\
    & &+\frac{N}{2}g^2 (1-\xi_{Q})^2 k^4 \int \frac{d^4 p}{(2\pi )^4}
     \frac{-p^{\mu}p^{\nu}}{p^4 q^4}.
\eea
where it is understood that the loop variables are related by $k + p + q =0$.

When we turn to the imaginary time finite temperature formulation,
we replace the integral in the Minkowski space with the following one:
\be
    \int \frac{d^4 p}{(2\pi )^4} \Rightarrow
                i\int \frac{d^3 p}{8\pi ^3} T \sum_n
\ee
where the summation goes over the $n$ in $p_0=2\pi inT$.
Applying the projection operator $t_{ij}$ to the spatial part of
$\Pi^{\mu \nu}_{P(BFG)}(k_0=0, \kappa=\vert \vec {\bf k} \vert)$,
we obtain in the limit $\kappa<<T$,
\bea
     \Pi_{T}^{P(BFM)}(T,\kappa)&=&N g^2 (1-\xi_{Q}) {\kappa}^2
 \int \frac{d^3 p}{8\pi ^3} T \sum_n \frac{2{\vec {\bf k}}\cdot {\vec {\bf p}}}
            {p^4 q^2}    \nonumber   \\
        & & - \frac{N}{4}g^2 (1-\xi_{Q})^2 {\kappa}^2
           \int \frac{d^3 p}{8\pi ^3} T \sum_n
        \frac{\vec {\bf k}^2 \vec {\bf p}^2 - ({\vec {\bf k}}.{\vec {\bf
p}})^2}
           {p^4 q^4} \\
&=&Ng^{2}{\kappa}T \biggl\{
              \frac{1}{8}(1-\xi_{Q})-
                    \frac{1}{64}(1-\xi_{Q})^2 \biggr\}  + {\cal O} (\kappa^2).
\label{PBFM}
\eea

Adding two contributions we find that the sum
\bea
\Pi_{T}(T,\kappa)&=&\Pi_{T}^{(BFM)}(T,\kappa)+\Pi_{T}^{P(BFM)}(T,\kappa)
     \nonumber  \\
       &=&Ng^{2}{\kappa}T \frac{7}{16}  + {\cal O} (\kappa^2)
\label{InvPi}
\eea
is  gauge-parameter $\xi_Q$ independent, and this gives a
gauge-independent thermmal $\beta$ function:
\be
      \beta_T=\frac{g^3 N}{128}\frac{7}{16}\frac{T}{\kappa}.
\label{InvBeta}
\ee
Note that the result coincides with $\beta_T^{BFM}$ in
Eq.(\ref{BetaBFM}) with $\xi_Q=1$~\cite{ACPS}.

\bigskip
\noindent
(ii) {\it The Feynman Gauge (The Covariant Gauge with $\xi=1$})

In the Feynman gauge (FG) the gluon propagator,
$iD_{ab(FG)}^{\mu \nu}=-i\delta _{ab}D_{(FG)}^{\mu \nu}$, is
in a very simple form:
\be
       D_{(FG)}^{\mu \nu}=\frac{1}{k^2}
                 g^{\mu \nu}.
\ee
The three-gluon vertex is expressed as
\be
\Gamma^{abc}_{\lambda \mu \nu } (p, k, q) = g f^{bac}
              \biggl[ \Gamma^{P}_{\lambda \mu \nu}(p, k, q)
              + \Gamma^{F}_{\lambda \mu \nu}(p, k, q) \biggr]
\label{VertexFG}
\ee
where $\Gamma^{P}_{\lambda \mu \nu}(p, k, q)$ and
$\Gamma^{F}_{\lambda \mu \nu}(p, k, q)$ are given in Eq.(\ref{Vertex}).
{}From the one-loop polarization tensor in FG, we extract
the contribution to the transverse function $\Pi_{T}^{(FG)}$
and find in the limit $\kappa << T$,
\be
   \Pi_{T}^{(FG)}(T,\kappa)=Ng^{2}{\kappa}T\frac{3}{16} + {\cal O} (\kappa^2).
\label{F}
\ee

The pinch contribution to the polarization tensor in FG  is
very simple. Since the gluon propagator in FG does not have a
$k^{\mu}k^{\nu}$ term, the only contribution is coming from
the vertex diagram of the second kind with the three-gluon
vertex $\Gamma^{P}$ (and its mirror graph)~\cite{rCP}, and it is given by
\be
  i\Pi^{\mu \nu}_{P(FG)} =2N g^2 k^2 \int \frac{d^4 p}{(2\pi )^4}
             \frac{1}{p^2 q^2} g^{\mu \nu}.
\ee
Following the same procedure as we did in (i), we obtain for the pinch
contribution to $\Pi_{T}^{FG}$ as
\be
     \Pi_{T}^{P(FG)}(T,\kappa) =
         Ng^{2}{\kappa}T\frac{1}{4} + {\cal O} (\kappa^2) .
\label{PFG}
\ee
Again when we add two contributions, $\Pi_{T}^{(FG)}$ and $\Pi_{T}^{P(FG)}$,
we obtain the same $\Pi_{T}$ in Eq.(\ref{InvPi}) and, thus, the same $\beta_T$
in Eq.(\ref{InvBeta}).

\bigskip
\noindent
(iii) {\it The Coulomb Gauge}

In the frame of a unit vector $n^{\mu}=(1,0,0,0)$,
the gluon propagator in the Coulomb gauge (CG),
$iD_{ab(CG)}^{\mu \nu}=-i\delta _{ab}D_{(CG)}^{\mu \nu}$, is defined by
\be
    D_{(CG)}^{\mu \nu}=\frac{P^{\mu \nu}}{k^2}+\frac{1}{k^2}
  \biggl[Q^{\mu \nu}+\frac{\sqrt2 k_0}{\vert {\vec {\bf k}} \vert}C^{\mu \nu}-
         \frac{k_0^2}{{\vec {\bf k}}^2}D^{\mu \nu}\biggr]
\ee
where
\bea
  P^{\mu \nu}(k) &=& g^{\mu \nu}-n^{\mu}n^{\nu}+
   \frac{1}{{\vec {\bf k}}^2}\biggl[k^{\mu}k^{\nu}-
     (n^{\mu}k^{\nu}+k^{\mu}n^{\nu})k_0+
       n^{\mu}n^{\nu}k_0^2 \biggr], \label{PP}     \\
   Q^{\mu \nu}(k) &=& -\frac{k^2}{{\vec {\bf k}}^2}n^{\mu}n^{\nu}+
     \frac{k_0}{{\vec {\bf k}}^2}\biggl[n^{\mu}k^{\nu}+
      k^{\mu}n^{\nu}\biggr]-
 \frac{k^{\mu}k^{\nu}k_0^2}{{\vec {\bf k}}^2 k^2}, \\
 C^{\mu \nu}(k)&=&-\frac{1}{\sqrt2 \vert {\vec {\bf k}} \vert}
        (n^{\mu}k^{\nu}+k^{\mu}n^{\nu})+
   \frac{\sqrt2 k_0}{\vert {\vec {\bf k}} \vert k^2}k^{\mu}k^{\nu}, \\
 D^{\mu \nu}(k)&=&
       \frac{k^{\mu}k^{\nu}}{k^2}.  \label{DD}
\eea
The three-gluon vertex is the same as in FG, that is,
$\Gamma^{abc}_{\lambda \mu \nu } (p, k, q)$ in Eq.(\ref{VertexFG}).

The transverse function $\Pi_T$ is related to
the polarization tensor as
\be
   \Pi_T=t_{ij}\Pi_{ij}=\frac{1}{2}\biggl[\Pi_{ii}-
           \frac{1}{\vec {\bf k}^2}k_i \Pi_{ij}k_j \biggr].
\ee
Since  $k_i \Pi_{ij}k_j=0$ in the static limit $k_0=0$, we have
$\Pi_T(T,\kappa)=\frac{1}{2}\Pi_{ii}(k_0=0, \kappa)$.
The $\Pi_{ii}$ for general $k_0$ and $\kappa$ was evaluated in CG
and the temporal axial gauge in Ref.~\cite{HKT}.
Using the expression of Eq.(4.38) in Ref.~\cite{HKT} for
$\Pi_{ii}^{CG}(k_0,\kappa)$, we can calculate the static limit of
$\Pi_{ii}^{CG}$ and obtain
\be
     \Pi_{T}^{(CG)}(T,\kappa)=\frac{1}{2}\Pi_{ii}^{(CG)}(0,\kappa)
               = Ng^{2}{\kappa}T\frac{9}{64} + {\cal O} (\kappa^2) .
\label{CG}
\ee

It is straightforward to calculate the pinch contributions to
the polarization tensor in CG. Again we consider the
quark-quark scattering at one-loop order. We discard the terms which
are proportional to $\nsl$ and $\ksl$,
where $k_{\mu}$ is the momentum transfered in the scattering,
and we make use of the Dirac equations satisfied by the external
quark fields, such as, $({\ppsl}-m) u(p)=0$ and
${\bar u(p)}({\ppsl}-m)=0$.
The pinch contribution to the polarization tensor in CG is expressed as
\bea
  i \Pi^{\mu \nu}_{P(CG)} &=& N g^2  k^2 \int \frac{d^4 p}{(2\pi )^4}
    \biggl[ \frac{k^2}{p^2 q^2 {\vec {\bf p}}^2} - \frac{1}{p^2
        {\vec {\bf p}}^2} -
 \frac{4 {\vec {\bf k}}\cdot {\vec {\bf p}}}{p^2 q^2 {\vec {\bf p}}^2}\biggr]
               g^{\mu \nu}  \nonumber \\
    & &+\frac{N}{2}g^2  k^2 \int \frac{d^4 p}{(2\pi )^4}
\frac{p_{\mu}p_{\nu}}{p^2 q^2}
\biggl[\frac{k^2}{{\vec {\bf p}}^2{\vec {\bf q}}^2} - \frac{4}{{\vec {\bf
p}}^2} +
         \frac{2{\vec {\bf k}}^2}{{\vec {\bf p}}^2{\vec {\bf q}}^2}\biggr].
\eea

Applying the projection operator $t_{ij}$ to the spatial part
$\Pi^{ij}_{P(CG)}$, we obtain in the static limit and for $\kappa <<T$
\bea
     \Pi_{T}^{P(CG)}(T,\kappa)&=&-N g^2 {\kappa}^2
 \int \frac{d^3 p}{8\pi ^3} T \sum_n \biggl\{ \frac{1}{q^2 \vec {\bf p}^2}+
   \frac{2{\vec {\bf k}}\cdot {\vec {\bf p}}}{p^2 q^2 \vec {\bf p}^2} \biggr\}
    \nonumber   \\
   & &- \frac{N}{4}g^2{\kappa}^2 \int \frac{d^3 p}{8\pi ^3} T \sum_n
    \biggl[1 - \frac{({\vec {\bf k}}\cdot {\vec {\bf p}})^2}
         {\vec {\bf k}^2 \vec {\bf p}^2}\biggr]
   \biggl\{\frac{\vec {\bf k}^2}{p^2 q^2 \vec {\bf q}^2}
            - \frac{4}{p^2 q^2} \biggr\} \\
   &=&Ng^{2}{\kappa}T\frac{19}{64} + {\cal O} (\kappa^2).
\label{PCG}
\eea

Adding the two contributions, $\Pi_{T}^{(CG)}$ and $\Pi_{T}^{P(CG)}$,
we find that the sum is equal to $\Pi_{T}$ in Eq.(\ref{InvPi}) and, thus,
we obtain the same $\beta_T$ in Eq.(\ref{InvBeta}).

\bigskip
\noindent
(iv) {\it The Temporal Axial Gauge}

The gluon propagator in the temporal axial gauge (TAG),
$iD_{ab(TAG)}^{\mu \nu}=-i\delta _{ab}D_{(TAG)}^{\mu \nu}$, is defined by
\be
    D_{(TAG)}^{\mu \nu}=\frac{P^{\mu \nu}}{k^2}+\frac{1}{k^2}
   \biggl[Q^{\mu \nu}+\frac{\vert {\vec {\bf k}} \vert}{\sqrt2 k_0}C^{\mu \nu}-
         \frac{{\vec {\bf k}}^2}{k_0^2}D^{\mu \nu}\biggr]
\ee
where $P^{\mu \nu}$, $Q^{\mu \nu}$, $C^{\mu \nu}$ and $D^{\mu \nu}$ are
given in Eqs.(\ref{PP}\ -\ref{DD}).  The three-gluon vertex is given by
$\Gamma^{abc}_{\lambda \mu \nu } (p, k, q)$ in Eq.(\ref{VertexFG}).
The static limit of $\Pi_{ii}^{(TAG)}$ was caluculated in Ref.~\cite{HKT}.
Using the result of Eq.(4.44) in Ref.~\cite{HKT}, we find
for $\kappa<<T$
\be
     \Pi_{T}^{(TAG)}(T,\kappa)=\frac{1}{2}\Pi_{ii}^{(TAG)}(0,\kappa)
               = Ng^{2}{\kappa}T\frac{5}{16}  + {\cal O} (\kappa^2).
\label{TAG}
\ee

Following the same procedure as we extracted the pinch parts from the
one-loop quark-quark scattering diagrams in CG,  we
obtain for the pinch contribution to the polarization tensor in TAG,
\bea
   i\Pi^{\mu \nu}_{P(TAG)} &=& N g^2  k^2 \int \frac{d^4 p}{(2\pi )^4}
          \biggl[ \frac{k^2}{p^2 q^2 p_0^2} - \frac{1}{p^2  p_0^2} +
        \frac{2}{q^2 p_0^2} -
     \frac{4 {\vec {\bf k}}\cdot {\vec {\bf p}}}{p^2 q^2 p_0^2}\biggr]
               g^{\mu \nu}  \nonumber \\
    & &+\frac{N}{2}g^2  k^2 \int \frac{d^4 p}{(2\pi )^4}
\frac{p_{\mu}p_{\nu}}{p^2 q^2}
     \biggl[\frac{k^2}{p_0^2 q_0^2} - \frac{4}{p_0^2} +
         \frac{2{\vec {\bf k}}^2}{p_0^2 q_0^2}\biggr].
\eea
Then in the static limit, $\Pi_{T}^{P(TAG)}$ is expressed as
\bea
     \Pi_{T}^{P(TAG)}(T,\kappa)&=& - N g^2 {\kappa}^2
   \int \frac{d^3 p}{8\pi ^3} T \sum_n \biggl\{
   \frac{{\vec {\bf k}}^2 + 4{\vec {\bf k}} \cdot {\vec {\bf p}}}
        {p^2 q^2 p_0^2} + \frac{1}{p^2  p_0^2} -  \frac{2}{q^2 p_0^2} \biggr\}
       \nonumber   \\
     & & - \frac{N}{4}g^2 {\kappa}^2 \int \frac{d^3 p}{8\pi ^3} T \sum_n
      \biggl[ \vec {\bf p}^2-\frac{(\vec {\bf k}\cdot \vec {\bf p})^2}
           {\vec {\bf k}^2} \biggr]
    \biggl\{ \frac{\vec {\bf k}^2}{p^2 q^2 p_0^2 q_0^2} -
        \frac{4}{p^2 q^2 p_0^2}  \biggr\}.
\label{Sing}
\eea
Due to the $1/{p_0^2}$ and $1/{q_0^2}$ terms coming from the
TAG propagator, the above integrand contains a $\vec {\bf k}^2 / \vec {\bf
p}^2$
singularity at the lower limit of the integration. This singularity is
circumvented by the principal value prescription~\cite{HKT}\cite{KUMMER}.
The result is for $\kappa<<T$
\be
   \Pi_{T}^{P(TAG)}(T,\kappa)=Ng^{2}{\kappa}T\frac{1}{8}+ {\cal O} (\kappa^2).
\label{PTAG}
\ee
Again the sum of $\Pi_{T}^{(TAG)}$ and $\Pi_{T}^{P(TAG)}$ coincides with
$\Pi_{T}$ in Eq.(\ref{InvPi}) and yields the same $\beta_T$ in
Eq.(\ref{InvBeta}).

I have demonstrated the calculation of
the thermal $\beta$ function $\beta_T$ in four
different gauges, that is, in BFM with an arbitrary gauge, in FG, in CG, and
in TAG. When the pinch contributions are taken care of, the same result
$\beta_T=\frac{g^3 N}{128}\frac{7}{16}\frac{T}{\kappa}$
was obtained at one-loop order in all four cases.
More details will be reported elsewhere~\cite{Sasakib}.
However, this is not the end of the story. Elmfors and Kobes
pointed out~\cite{EK} that the leading contribution to $\beta_T$,
which gives a term
$T/\kappa$,  does not come from
the hard part of the loop integral, responsible  for a $T^2/\kappa^2$ term,
but from soft loop integral. Hence they emphasized that it is not consistent
to stop the calculation at one-loop order for soft internal momenta and that
the resummed propagator and vertices~\cite{BP} must be used to get the
complete leading contribution. Since the corrections to the bare
propagator and vertices, which come from the hard thermal loops,
are gauge-independent and satisfy simple Ward identities~\cite{BP},
it is well-expected that we will obtain the gauge-independent thermal
$\beta$ function even when we use the resummed propagator and vertices
in the framework of PT. Study along this direction is under way.

\bigskip
\bigskip

I would like to thank Rob Pisarski for the hospitality extended to me at
Brookhaven National Laboratory in the summer of 1995. I was inspired to
start this work by the ``hot" atmosphere created  at the laboratory.
I would also like to thank A. Sirlin for the hospitality extended to me
at New York University where this work was completed. Finally I am very
grateful to J. Papavassiliou for helpful discussions.
This work is partially supported by Yokohama National University
Foundation.


\bigskip
\medskip
\noindent
{\large\bf Figure caption}
\medskip

\noindent
Fig.1

\noindent
The self-energy diagram for the quark-quark scattering.

\medskip

\noindent
Fig.2

\noindent
(a) The vertex diagrams of the first kind for
the quark-quark scattering.
(b) Their pinch contribution.

\medskip

\noindent
Fig.3

\noindent
(a) The vertex diagram of the second kind
for the quark-quark scattering.
(b) Its pinch contribution.

\medskip

\noindent
Fig.4

\noindent
(a) The box diagrams for the quark-quark scattering.
(b) Their pinch contribution.


\newpage
\begin{thebibliography}{99}
\bibitem{Umezawa}
     H. Matsumoto, Y. Nakano, and H. Umezawa, {\sl Phys. Rev.} {\bf D29},
         1116 (1984).
\bibitem{EK}
     P. Elmfors and R. Kobes,
     {\sl Phys. Rev.} {\bf D51}, 774 (1995).
\bibitem{Niegawa}
     H. Nakkagawa, A. Ni$\acute {\rm e}$gawa, and H. Yokota,
       {\sl Phys. Rev.} {\bf D38}, 2566 (1988).


\bibitem{BFM}
   N. P. Landsman, {\sl Phys. Lett.} {\bf B232}, 240 (1989);
   M. A. van Eijck, {\sl Can. J. Phys.} {\bf 71}, 237 (1993);
   M. A. van Eijck, C. R. Stephens, and Ch. G. van Weert,
         {\sl Mod. Phys. Lett.} {\bf A9}, 309 (1994).
\bibitem{ACPS}
   J. Antikainen, M. Chaichian, N. R. Pantoja, and J. J. Salazar,
    {\sl Phys. Lett.} {\bf B242}, 412 (1990).
\bibitem{Dea}
     B. S. DeWitt, {\sl Phys. Rev.} {\bf 162}, 1195 (1967); also in
     {\sl Dynamical Theory of Groups and Fields}
     (Gordon and Breach, New York, 1963).
\bibitem{EHKT}
     H.-Th. Elze, U. Heinz, K. Kajantie, and T. Toimela,
     {\sl Z. Phys.} {\bf C37}, 305 (1988).

\bibitem{Papavass}
    J. Papavassiliou, {\sl Phys. Rev.} {\bf D51}, 856 (1995).
\bibitem{SirlinKostas}
    K. Philippides and A. Sirlin, {\sl Nucl. Phys.} {\bf B450}, 3 (1995).

\bibitem{rCa}
     J. M. Cornwall, in {\sl Proceedings of the French-American Seminar
     on Theoretical Aspects of Quantum Chromodynamics},
     Marseille, France, 1981, edited by J. W. Dash
     (Centre de Physique Th\'eorique, Marseille, 1982);
     {\sl Phys. Rev.} {\bf D26}, 1453 (1982).
\bibitem{rCP}
     J. M. Cornwall and J. Papavassiliou, {\sl Phys. Rev.} {\bf D40},
     3474 (1989); J. Papavassiliou, {\sl Phys. Rev.} {\bf D47}, 4728 (1993).
\bibitem{StandardModel}
     J. Papavassiliou, {\sl Phys. Rev.} {\bf D41}, 3179 (1990);
     G. Degrassi and A. Sirlin, {\sl Phys. Rev.} {\bf D46}, 3104 (1992);
     J. Papavassiliou and K. Philippides,
             {\sl Phys. Rev.} {\bf D48}, 4255 (1993);{\bf D52}, 2355 (1995);
     G. Degrassi, B. Kniehl, and A. Sirlin,
          {\sl Phys. Rev.} {\bf D48}, R3963 (1993);
     J. Papavassiliou and C. Parrinello,
             {\sl Phys. Rev.} {\bf D50}, 3059 (1994);
     J. Papavassiliou and A. Sirlin,
             {\sl Phys. Rev.} {\bf D50}, 5951 (1994);
     J. Papavassiliou, {\sl Phys. Rev.} {\bf D50}, 5958 (1994);
     K. Hagiwara, D. Haidt, C. S. Kim, and S. Matsumoto,
          {\sl Z. Phys.} {\bf C64}, 559 (1994);
     J. Papavassiliou, K. Philippides, and M. Schaden,
             {\sl Phys. Rev.} {\bf D51}, 6364 (1995);
     J. Papavassiliou and A. Pilaftsis,
             hep-ph/9506417; hep-ph/9507246;
     J. Papavassiliou, K. Philippides, and K. Sasaki, hep-ph/9507265

\bibitem{Hiroshima}
    A. Denner, S. Dittmaier, and G. Weiglein,
           {\sl Phys. Lett.} {\bf B333}, 420 (1994);
           {\sl Nucl. Phys.} {\bf B440}, 95 (1995);
    S. Hashimoto, J. Kodaira, K. Sasaki, and Y. Yasui,
           {\sl Phys. Rev} {\bf D50}, 7066 (1994);
    E. de Rafael and N. J. Watson, (unpublished).






\bibitem{Abbott}
     L. F. Abbott,
     {\sl Nucl. Phys.} {\bf B185}, 189 (1981);
     {\sl Acta Physica Polonica} {\bf B13}, 33 (1982).
\bibitem{HKT}
     U. Heinz, K. Kajantie, and T. Toimela,
     {\sl Annals of Physics} {\bf 176}, 218 (1987).
\bibitem{KUMMER}
   W. Kummer, {\sl Acta Phys. Austriaca} {\bf 41}, 315 (1975).
\bibitem{Sasakib}
     K. Sasaki, in preparation.
\bibitem{BP}
     E. Braaten and R. D. Pisarski,
    {\sl Nucl. Phys.} {\bf B337}, 569 (1990); {\bf B339}, 310 (1990)



\end{thebibliography}
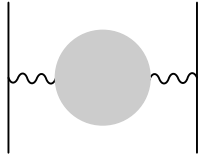
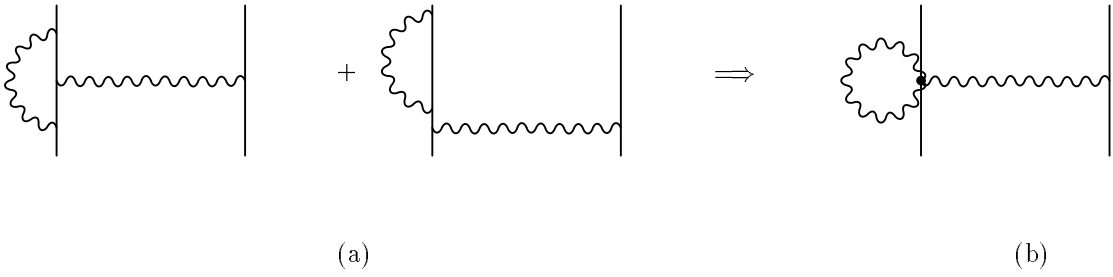
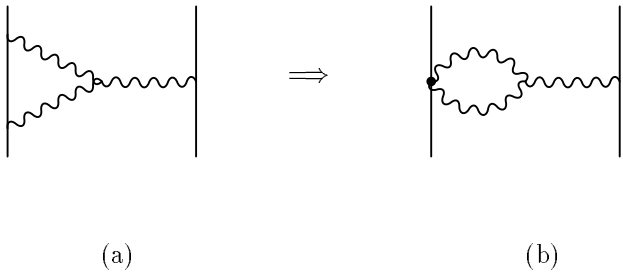
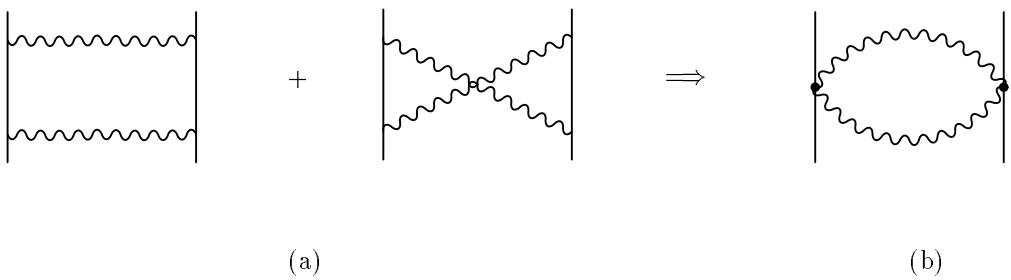
\end{document}